\documentclass[aps,preprint,showpacs,superscriptaddress,groupedaddress]{revtex4}  
\usepackage{graphicx}
\usepackage{dcolumn}   
\usepackage{bm}        
\usepackage{amssymb}   
\usepackage{amsmath}
\usepackage{url,color}
\usepackage{layout}
\usepackage{epsfig}
\usepackage{graphicx}
\usepackage{epstopdf}
\usepackage{booktabs}
\usepackage{float,CJK}
\usepackage{array}
\usepackage{graphicx}
\usepackage{textpos}

\usepackage[hyperindex,breaklinks]{hyperref}
\newcommand{\ket}[1]{|#1\rangle}                  
\newcommand{\inner}[2]{\langle #1|#2\rangle}

\begin{document}
\title{Resonant Quantum Search with Monitor Qubits}
\begin{CJK}{UTF8}{gbsn}
\author{Frank Wilczek}
\affiliation{Center for Theoretical Physics, MIT, Cambridge MA 02139 USA}
\affiliation{T. D. Lee Institute, Shanghai Jiao Tong University, Shanghai 200240, China}
\affiliation{Wilczek Quantum Center, School of Physics and Astronomy,
Shanghai Jiao Tong University, Shanghai 200240, China}
\affiliation{Department of Physics, Stockholm University, Stockholm SE-106 91 Sweden}
\affiliation{Department of Physics, Arizona State University, Tempe AZ 25287 USA}
\author{Hong-Ye Hu(扈鸿业)}
\affiliation{Department of  Physics, University of Californian, San Diego, CA 92093, USA}
\author{Biao Wu(吴飙)}
\affiliation{International Center for Quantum Materials, School of Physics, Peking University, 100871, Beijing, China}
\affiliation{Wilczek Quantum Center, School of Physics and Astronomy, Shanghai Jiao Tong University, Shanghai 200240, China}
\affiliation{Collaborative Innovation Center of Quantum Matter, Beijing 100871,  China}

\preprint{MIT-CTP-5177}

\begin{abstract}
We present an algorithm for the generalized search problem (searching $k$ marked items among $N$ items) based on a continuous Hamiltonian and exploiting resonance.  This resonant algorithm has the same time complexity $O(\sqrt{N/k})$ as 
the Grover algorithm. A natural extension of the algorithm, incorporating auxiliary ``monitor'' qubits, can determine $k$ precisely, if it is unknown. The time complexity
of our counting algorithm is  $O(\sqrt{N})$, similar to the best quantum approximate counting algorithm, or better, given appropriate physical resources. 
\end{abstract}

\begin{textblock*}{5cm}(11cm,-8.2cm)
\fbox\footnotesize MIT-CTP/5177
\end{textblock*}

\maketitle
\end{CJK}

\section{Introduction} 
The possible advantages of quantum computers over classical computers are rooted in 
the tensor product structure of quantum mechanics and the superposition principle. It is, however, not straightforward to 
utilize those advantage. 
Shor's algorithm for factorizing large integer numbers~\cite{Shor} 
and Grover's search algorithm~\cite{Grover} are outstanding but rare examples of presently known cases where quantum computation gives a theoretical edge in  
a natural problem ~\cite{ChuangBook}.  Shor's  and Grover's  
algorithms are quantum circuit algorithms, 
consisting of a sequence of discrete operations known as quantum gates~\cite{ChuangBook}. 

There is a different paradigm of quantum computing
wherein algorithms are designed by constructing Hamiltonians.  The system is initially in an easy-to-prepare quantum state, 
and the quantum computer evolves the quantum state using designed Hamiltonians. It eventually arrives at a quantum state which
encodes the solution of the problem. The Hamiltonian approach can take advantage 
of intuition in quantum mechanics that physicists have cultivated over decades of research. 
A Hamiltonian was proposed for quantum search 
by Farhi and Gutmann in 1998~\cite{Farhi1998}, and a generic quantum adiabatic 
algorithm was proposed in 2000~\cite{Farhi2000}. In the adiabatic algorithm, the quantum computer follows the ground state of a time-dependent  Hamiltonian.   
It has been shown that every 
quantum circuit algorithm can be converted into a quantum adiabatic algorithm, 
whose time complexity is exactly the same~\cite{Dam,YHW}. A quantum Hamiltonian algorithm for independent-set problems has some advantages over other known quantum algorithms and
classical algorithms~\cite{WYW}.

As previously mentioned, a quantum algorithm is essentially a manipulation that evolves an initial state to a target quantum state.  Since   
resonance has been widely exploited in many branches of physics to achieve that sort of state evolution, 
it is natural to ask whether resonant evolution might be useful in this context. 

Here we use resonance to construct a quantum Hamiltonian algorithm for a generalized form of the problem addressed by Grover, namely to find, given an oracle, marked entries within a list of items.  If the list has $N$ entries, and there are $k \geq 1$ marked items, our resonant algorithm can find one of the marked entries in time $O(\sqrt{N/k})$.   This time complexity is the same as the Grover algorithm~\cite{Grover}
and the quantum adiabatic search algorithm~\cite{Dam1,Cerf2000}. Though there no gain in performance, there is no loss either, and the resonant approach seems particularly simple and transparent.  

Next we introduce the concept of a monitor qubit, which is very natural in our context.  Roughly speaking, a monitor qubit keeps track of whether the resonant transition of interest has occurred.  Through use of monitor bits we can both avoid wasteful measurements on computational bits, and also gather information on the initial state.  Below we demonstrate two different, characteristic methods to extract information using monitor qubits: predictive dissonance and robust readout.  In the search context, predictive dissonance allows us to determine the number $k$ of marked entries, when it is not given, with the time complexity $O(\sqrt{N})$. All known quantum algorithms can only approximately determine $k$  with a similar time complexity~\cite{Wie,Rall}.   Robust readout is a more open-ended concept, which depends in detail on the physical implementation of the quantum computer.   Given appropriate resources, it can speed things up further.


\section{Quantum Resonance  Search Algorithm}
Let us briefly recall the basic resonance phenomenon in a two-state quantum 
system~\cite{Scully}.  We consider the time-dependent Hamiltonian,
\begin{equation}
	\hat{H}(t) = \begin{pmatrix}
		\dfrac{\Delta}{2} & \epsilon e^{-i\omega t}\\
		
		\epsilon e^{i\omega t} & -\dfrac{\Delta}{2}
	\end{pmatrix}\,.
	\label{two}
\end{equation}
where $\Delta$ is the energy difference between the two states and $\omega$ and $\epsilon$ are the 
frequency and strength of the external drive, respectively. Without loss of generality, 
we assume that $\epsilon$ is real.  This Hamiltonian
can describe some realistic physical systems, or arise as a rotating-wave approximation
of systems where the driving is proportional to $\cos (\omega t)$.  
By going to a rotating reference frame in Hilbert space, one readily derives the time evolution operator corresponding to Eq.\,(\ref{two}) to be

\begin{equation}
	\hat{U}(t) = \cos(\kappa t)\hat{I}-i\sin(\kappa t)\Big[\dfrac{\omega-\Delta}{2\kappa}\hat{\sigma}_{z}+\dfrac{\epsilon}{\kappa}\hat{\sigma}_{x}\Big]\,,
\end{equation}
where $\kappa= \sqrt{\epsilon^2+(\omega-\Delta)^2/4}$. 

Off resonance, when
$|\omega-\Delta|\gg |\epsilon|$\,, we have
\begin{equation}
	\hat{U}(t)\approx \cos(\dfrac{|\omega-\Delta|}{2}t)\hat{I}-i\sin(\dfrac{|\omega-\Delta|}{2}t)\hat{\sigma}_{z}.
\end{equation}
In this case, if the initial condition has only upper component, then the system will remain concentrated on the upper component forever. 

On resonance, $|\omega-\Delta|\ll\epsilon $, the time propagator becomes
\begin{equation}
	\hat{U}(t) \approx \cos(\epsilon t)\hat{I}-i\sin(\epsilon t)\hat{\sigma}_{x}\,.
\end{equation}
If we start with only the upper component present, it will have evolved completely into a state with only the lower component after a time $\tau = \pi/(2\epsilon)$. 

We now apply this framework to construct a quantum search algorithm.  The basic 
search problem is to find items that satisfy certain criteria in an unsorted database that contains $N$ items. On a quantum computer, these items are stored as $n = \log_{2}N$ qubits with $N$ orthonormal basis states $\ket{1}, \ket{2}, \cdots, \ket{N}$ embodying a binary encoding. 
To exploit quantum resonant search, we construct the Hamiltonian
\begin{equation}
	\hat{H}(t) = a(t)\hat{H}_{\gamma}+b(t)\hat{H}_{x}+c(t),
	\label{resonance}
\end{equation}
where $\hat{H}_{\gamma} = |\gamma\rangle\langle \gamma |$ and $\hat{H}_{x} = |x\rangle\langle x|$. 
The state $\ket{\gamma} =\dfrac{1}{\sqrt{N}} \sum_{j}\ket{j}$ is the equal-weight superposition of the number basis. Since $\ket{x}$ is the state that satisfies our searching criteria, we call it the answer state. In general, there could be more than one state that satisfy the searching criteria, and we will discuss those scenarios shortly.  $\hat{H}_{x}$ embodies the oracle~\cite{Dam1,Cerf2000} which encodes the answer.

As the initial state and the Hamiltonian have
the same permutation symmetry, we decompose the quantum state
 $|\psi\rangle$ as
\begin{equation}
	|\psi\rangle =\phi_1\dfrac{\sqrt{N-1}}{\sqrt{N}}|x_{\perp}\rangle+
	\phi_2\dfrac{1}{\sqrt{N}}|x\rangle
	\label{reduce}
\end{equation}
Here $|x_{\perp}\rangle=\dfrac{1}{\sqrt{N-1}} \sum^\prime_{j}\ket{j}$, where the summation is over all items 
other than the answer item. This converts the system into a two-state  model spanned by $|x\rangle$ and $|x_{\perp}\rangle$. 
The Hamiltonian (\ref{resonance}) now takes the reduced form
\begin{equation}
	\hat{\mathcal H}_1(t)=\begin{pmatrix}
		a(t)+c(t) & \sqrt{\frac{1}{N}}a(t)\\
		\\
		\sqrt{\frac{1}{N}}a(t) &  b(t)+c(t)
	\end{pmatrix}\,,
\end{equation}
where we have taken the large $N$ limit.  Now we choose
\begin{eqnarray}
		a(t) &=& p\cos(\omega t)\,,\\ 
		b(t) &=& -\Delta+p\cos(\omega t)\,,\\ 
		c(t) &=& \Delta/2-p\cos(\omega t)\,.
\end{eqnarray}
By comparing it to the Hamiltonian in Eq.(\ref{two}), 
we have $\epsilon=p/(2\sqrt{N})$.  (Here we have invoked the rotating wave approximation.  It could be avoided, as before, with a slightly different, notationally more complicated Hamiltonian.)
As our initial state is mostly in the upper component, 
$\inner{s}{x_\perp}\approx 1$, we see that it will have rotated to the desired item $\ket{x}$ 
after $\tau_1=\pi\sqrt{N}/p$.  If $p$ is independent of $N$, the time complexity of our algorithm is $O(\sqrt{N})$, 
the same as the Grover algorithm~\cite{Grover} and the quantum adiabatic search algorithm~\cite{Dam1,Cerf2000}. 

Note that $a(t) = 1-\dfrac{t}{T}, b(t) = -c(t)=-\dfrac{t}{T}$ corresponds to the adiabatic quantum search Hamiltonian 
of Farhi and Gutmann~\cite{Farhi1998}. 

A simple variation on the basic search problem is to allow $k$ different valid answers. 
Similarly, we can decompose the Hilbert space into two: one spanned by the $k$ answer items that spanned sub-space $\mathcal{M}$,
and the rest space spanned by the other items.  As long as $k\ll N$,  we have 
in the large $N$ limit 
\begin{equation}
	\hat{\mathcal H}_1(t)=\begin{pmatrix}
		a(t)+c(t) & \sqrt{\frac{k}{N}}a(t)\\
		\\
		\sqrt{\frac{k}{N}}a(t) &  b(t)+c(t)
	\end{pmatrix}\,,
\end{equation}
The critical rotation time is then $\tau_k= \pi\sqrt{N/k}/p$.

\section{Monitor Qubits}

We define a monitor qubit by expanding the Hilbert space to include an auxiliary qubit (i.e., the monitor qubit) and generalizing $a(t)$, $b(t)$, and $c(t)$ in the form
\begin{eqnarray}
\hat{a}(t) &=& \hat{1}\otimes \hat{\sigma}_{x}p\cos(\omega t)\,,\\ 
\hat{b}(t) &=& \hat{1}\otimes \hat{\sigma}_{x}p\cos(\omega t)-\Delta \hat{1}\otimes \hat{1}\,,\\ 
\hat{c}(t) &=& \dfrac{\Delta}{2} \hat{1}\otimes \hat{1}-\hat{1}\otimes\hat{\sigma}_{x}p\cos(\omega t)\,. 
\end{eqnarray}
where of course the second factor acts on the monitor qubit. 
We again use the rotating wave approximation 
and Eq.(\ref{reduce}) to reduce the Hamiltonian.  In the rotating frame, we have 
\begin{equation}
	\hat{H}_{rot} = \big(\dfrac{\omega-\Delta}{2}\big)\hat{\sigma}_{z}\otimes \hat{1}
	+\epsilon\hat{\sigma}_{x}\otimes \hat{\sigma}_{x}\,.
\end{equation}  
On resonance $|\omega-\Delta|\ll \epsilon$ the time evolution operator is 
\begin{equation}
	\hat{U}(t) \approx \cos(\epsilon t)\hat{1}\otimes\hat{1}-i\sin(\epsilon t)
	\hat{\sigma}_{x}\otimes \hat{\sigma}_{x}\,,\label{eqn:evolution}
\end{equation}
demonstrating that the monitor qubit rotates simultaneously with the computational qubits. 

If the initial state is prepared to be 
\begin{equation}
\label{psi0}
\ket{\psi(0)}=\left(\dfrac{1}{\sqrt{N}}\ket{x}+\dfrac{\sqrt{N-1}}{\sqrt{N}}\ket{x_\perp} \right)\otimes \ket{0}
\end{equation}
then following the dynamics given by Eqn.\,(\ref{eqn:evolution}), we find
\begin{eqnarray}
&&\ket{\psi(t)}=\cos(\epsilon t)\left(\dfrac{1}{\sqrt{N}}\ket{x}+\dfrac{\sqrt{N-1}}{\sqrt{N}}\ket{x_\perp} \right)\otimes |0\rangle \nonumber\\
&&~~~- i\sin(\epsilon t)\left(\dfrac{1}{\sqrt{N}}\ket{x_\perp}+\dfrac{\sqrt{N-1}}{\sqrt{N}}\ket{x} \right)\otimes |1\rangle.
\end{eqnarray}
This allows us to make measurement on the monitor qubit without collapsing the computational qubits to their number states.  
Suppose we make a measurement at time $t$ on the monitor qubit. If the result is $\ket{1}$, the system collapses 
to $\left(\frac{1}{\sqrt{N}}\ket{x_\perp}+\frac{\sqrt{N-1}}{\sqrt{N}}\ket{x} \right)\otimes \ket{1}$. In the case, 
we measure the computational qubits and will find the answer with probability $(N-1)/N$. 
If the result  is $\ket{0}$,  the  system will collapse to state 
$\left(\frac{1}{\sqrt{N}}\ket{x}+\frac{\sqrt{N-1}}{\sqrt{N}}\ket{x_\perp} \right)\otimes \ket{0}$, 
which is exactly the initial state $\ket{\psi(0)}$ we prepared. Therefore we can continue to run the algorithm without the need to
re-initialize the system.  This could be useful, in the case $k$ is known, if we have small errors which take us off exact resonance and introduce rare failures.

More interesting is the possibility to address the general problem of determining $k$, when it is not given.  
This has been known as quantum counting problem~\cite{Wie,Rall}. We will discuss two approaches to that problem.  The first involves the concept of predictive dissonance.   The second involve the concept of robust readout.  Both of those concepts are of independent interest.  They are characteristic potentialities opened up by monitor qubits, and could be of wider utility.   

\section{Predictive Dissonance} In Eq.\,(\ref{eqn:evolution}) we must take
\begin{equation}
\epsilon ~=~ \frac{p\sqrt k}{2 \sqrt N}
\end{equation}
As a consequence, there will be times 
\begin{equation}
t^{\rm zero} (k) ~=~ l \pi  \frac{2 \sqrt N}{p\sqrt k}, 
\end{equation}
where $l$ is an integer, when the monitor qubit (initially $| 0 \rangle$) is surely 0 and times
\begin{equation}
t^{\rm one} (k) ~=~ (l + \frac{1}{2}) \pi  \frac{2 \sqrt N}{p\sqrt k}
\end{equation}
when the monitor qubit is surely 1.   The case $k = 0$ is special: then the monitor qubit is always 0.  

Now given values $k_1, k_2$, we would like to find times for which $k_1$ predicts the monitor qubit to be 0 and $k_2$ predicts it to be 1, or {\it vice versa}, i.e.
\begin{eqnarray}\label{case1}
t ~&=&~ 2 l_1  \pi \frac{\sqrt N}{p\sqrt k_1} \nonumber \\
~&=&~ (2  l_2 + 1)  \pi  \frac{ \sqrt N}{p\sqrt k_2}
\end{eqnarray}
or 
\begin{eqnarray}\label{case2}
t ~&=&~ (2l_1 + 1) \pi  \frac{\sqrt N}{p\sqrt k_1} \nonumber \\
~&=&~  2 l_2  \pi  \frac{\sqrt N}{p\sqrt k_2}
\end{eqnarray}
for integers $l_1, l_2$.  We will refer to this phenomenon where alternative hypotheses give contradictory predictions, exactly or with high probability, as ``predictive dissonance".  In our context, it is related to the physical phenomenon of beats.  Predictive dissonance is a way to insure progress.  By measuring the monitor qubit at such a time, we will rule out either $k_1$ or $k_2$. For example, in the case of Eq.(\ref{case1}), if the monitor bit is measured to be 0, $k_2$ can be ruled out; 
if the monitor bit is 1, $k_1$ can be ruled out.  
And thus, if we are given an upper bound $k_{\rm max}$ on the possible values of $k$, we can home in a unique $k$ 
after at most $k_{\rm max}$ invocations of predictive dissonance. Our numerical results show that 
the number of invocations is proportional to $k_{\rm max}^\alpha$ with $\alpha\lesssim 0.7$ (see next section).

Unfortunately it is not always possible to achieve exact predictive dissonance.  For one thing, the occurrence of square roots of $k_1$ and $k_2$ in generally precludes the existence of such times.    On the other hand, by careful consideration of  
$\sqrt{k_1/k_2}$ we can find times which satisfy our requirements to a good approximation. 
At such times, we can interpret the measurement of the monitor qubit as ruling out $k_1$ or $k_2$ with high probability.   Of course, for efficiency we also want to keep the times reasonably small.   

We can assume that $k_1 < k_2$,  First suppose that $\sqrt {\frac{k_2}{k_1}}$ is rational, and write it in the reduced form $2^s\frac{a}{b}$ with $a, b$ odd.  Then if $s < 0$ we can satisfy Eq.\,(\ref{case1}) with
\begin{eqnarray}\label{result1}
l_1 ~&=&~ 2^{-s-1} b \nonumber \\
2l_2 + 1 ~&=&~ a \nonumber \\
t ~&=&~ \frac{2^{-s} \pi \sqrt N}{p \sqrt {k_1}} ~\leq~ \frac{\pi \sqrt N}{p}
\end{eqnarray}
while if $s>0$ we can satisfy Eqn.\,(\ref{case2}) with 
\begin{eqnarray}\label{result2}
2l_1 + 1 ~&=&~ b \nonumber \\
l_2 ~&=&~ 2^{s-1} a \nonumber \\
t ~&=&~ \frac{2^{s-1} \pi \sqrt N}{p \sqrt {k_2}} ~\leq~ \frac{\pi \sqrt N}{p}
\end{eqnarray}
In the exceptional case $s=0$ we do not get exact predictive dissonance, but we can get close, as follows.  At times $t = 2 l_2  \pi  \frac{\sqrt N}{p\sqrt {k_2}}$ we will surely measure 0 if $k = k_2$ on the monitor qubit, while if $k = k_1$ we will measure 1 with probability
\begin{equation}
P_1 ~\equiv~ \sin^2 ( l_2 \pi \sqrt {\frac{k_1}{k_2}} ) ~=~ \sin^2 (\pi l_2 \frac{b}{a}) 
\end{equation}
Now elementary number theory instructs us that there will be values of $l_2 < a $ for which
\begin{equation}
l_2 b ~\equiv~ \frac{a \pm 1}{2} \ \ \ ({\rm mod} \ a)
\end{equation}
For these values of $l_2$ we will have
\begin{equation}\label{p1_result}
P_1 =  \cos^2 \frac{\pi}{2a} \geq .75
\end{equation}
since $a \geq 3$.   Thus if we measure 1 we can eliminate $k_2$ as a candidate, while if we measure 0 repeatedly we can eliminate $k_1$ with high confidence.  For each measurement, the same time bound we saw in Eqs.\,(\ref{result1}, \ref{result2}) applies.

We now switch to a different procedure, cruder but more general, which applies  when $\sqrt{\frac{k_2}{k_1}}$ is irrational.  (Number-theoretic refinements are certainly possible, but they are beyond the scope of this paper.). To set the stage, let us re-state the essence of our problem in the form we will address it.   We want to set up predictive dissonance by finding a time, not too large, such that on resonance measurement of the monitor qubit will surely yield 0 if $k= k_2$ but will have large probability to yield 1 if $k=k_1$.  The first condition reads
\begin{eqnarray}
{\rm phase}_2 ~&=&~ \frac{2p\sqrt{k_2}}{\sqrt N} t  ~=~ l_2 \pi \nonumber \\
t ~&=&~ \frac{l_2 \pi \sqrt N}{2p\sqrt{k_2}}
\end{eqnarray}
and gives us
\begin{equation}
{\rm phase}_1 ~=~ {l_2 \pi} \sqrt{\frac{k_1}{k_2}}
\end{equation}
We want to insure that ${\rm phase}_1$ is close to $\frac{\pi}{2}$ modulo $\pi$, and also, in order for our time bound to hold, that $l_2 \leq \sqrt{k_2}$.   Let us consider the phase modulo $\pi$ as defining a circle.   If $\pi \sqrt{\frac{k_1}{k_2}}$ lies within the interval of length $\frac{\pi}{3}$ centered at $\frac{\pi}{2}$, then simply by choosing $l_2 =1$ we achieve
\begin{equation}
P_1 ~\geq~ \cos^2 \frac{\pi}{6} ~=~ .75
\end{equation}
as in Eq.\,(\ref{p1_result}).  If $\theta \equiv \pi \sqrt{\frac{k_1}{k_2}}$ lies in the interval $0 < \theta \leq \frac{1}{3} \pi$, modulo $\pi$, then steps in units of $\theta$ will move us monotonically into the sector just described.  If $\theta$ lies in the interval $ \frac{2}{3} \pi \leq \theta < \pi $, then steps in units of $\theta$ will move us monotonically backward into that sector.    One can check that the number of steps required is always consistent with our standard time bound.  Finally, the case $\theta = 0$, corresponding to $k_1 =0$, is trivial. 

\section{Numerical simulation with predictive dissonance}
We now apply predictive dissonance to a class of problems, where we have an estimation of the maximum number 
of possible solutions $k_{\text{max}}$ with $k_{\text{max}}$ being independent of $N$. For many hard instances of NP complete 
problems, this is indeed the case~\cite{Farhi2001}. We want to pinpoint the number of solutions, $k_{\text{true}}\in [0,k_{\text{max}}]$. We can choose pairs of $k_1$ and $k_2$ in the range $[0,k_{\text{max}}]$, and use predictive dissonance to eliminate one of them after the readout. In general, the choice of $k_1$ and $k_2$ will result in $\sqrt{\frac{k_2}{k_1}}$ as an irrational number. Then we could use the protocol described in the previous section to choose the proper time $t_{\text{run}}$ such that the measurement of monitor qubit will surely yield 0 if $k=k_2$ and will have high probability $p$ to yield 1 if $k=k_1$. In fact, the protocol described in the previous section ensures $p\geq 0.75$. To further enhance the probability $p$, we take a sequential $J$ measurements of monitor qubit, and the readout will be a binary string of length $J$, i.e. $R = [0,0,1,0,\cdots]$, where $0$ means no flip of the monitor qubit and $1$ denotes the flip of the monitor qubit. If there is at least one $1$ in the readout $R$, we can eliminate $k_{2}$. If the readout $R$ has only $0$, then we can eliminate $k_1$ confidently, because the probability of such a case appearing is $(1-P)^{J}\ll 1$. The general time complexity of our predictive dissonance protocol can be expressed as $O(k_{\text{max}}^{\alpha}N^{\beta})$. We expect $\beta=0.5$ because the single run time $t_{\text{run}}\propto \sqrt{N}$. As will be shown below, $\alpha$ depends on the detail of choosing $k_1$, $k_2$ pairs. In the following, we discuss two pairing schemes: (1) half-size pairing and (2) head-tail pairing.

\begin{figure}
    \centering
    \includegraphics[width = 0.7\linewidth]{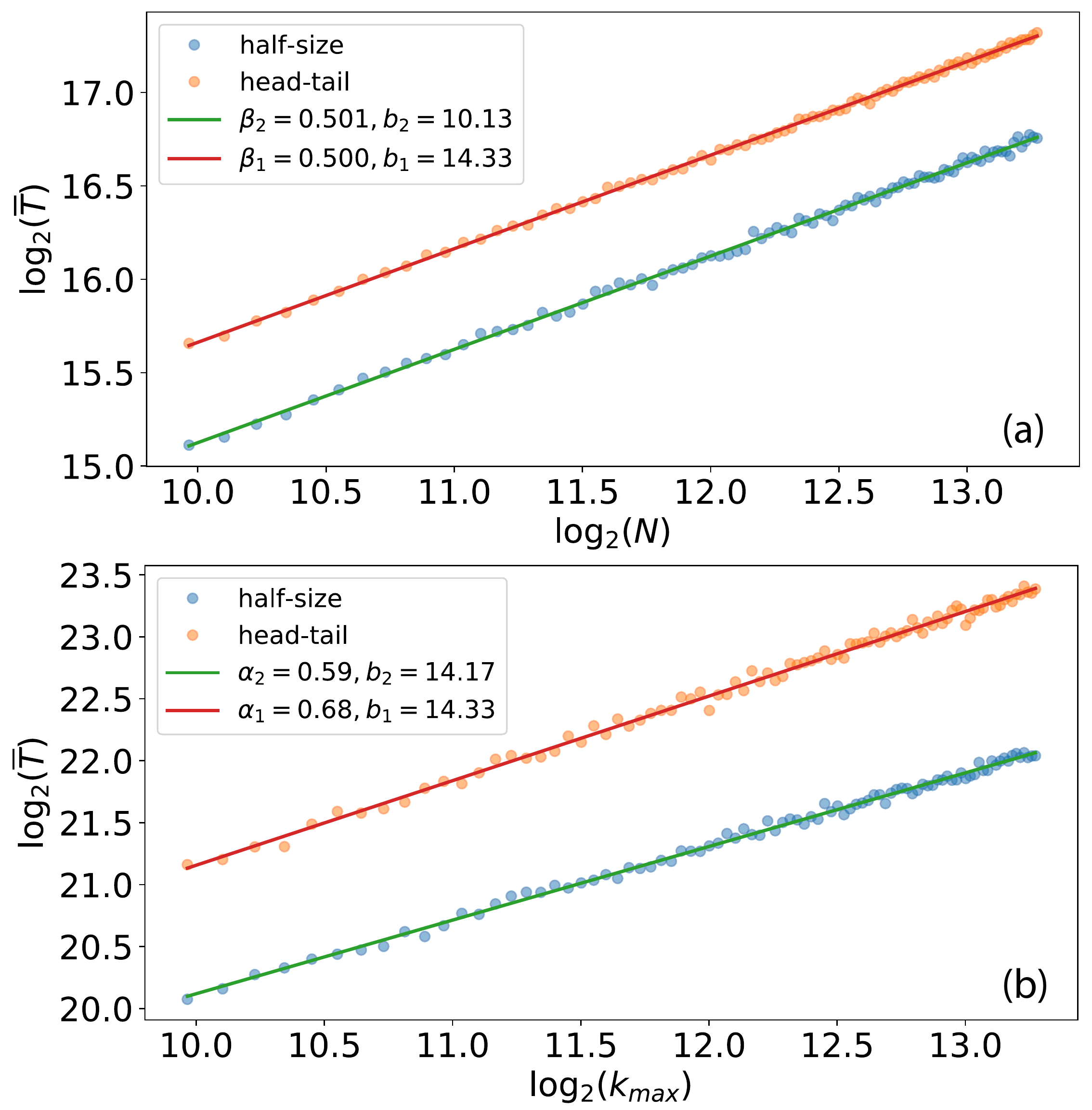}
    \caption{(color online)Scaling behavior of ensemble averaged running time $\overline{T}$ as a function of $N$ and $k_{\text{max}}$. For a given $N$ and $k_{\text{max}}$, we uniformly sample $k_{\rm true}\in [0,k_{\text{max}}]$ 300 times. For each $k$ sample, we follow the predictive dissonance protocol to pinpoint $k$ and record the running time $T$, and we choose repetition $J=6$. The ensemble averaged time $\overline{T}$ is plotted for each $N$ and $k_{\text{max}}$. In subplot (a), it shows $\overline{T}\propto N^{\beta}$, where $\beta$ is approximately 0.5; in subplot (b), it shows $\overline{T}\propto k_{\text{max}}^{\alpha}$, and $\alpha$ depends on the details of pairing schemes.}
    \label{fig:scaling}
\end{figure}

At a given time, we always have a list of potential $\{k_{i}\}$ and we list them in an increasing order: $k_0<k_1<\cdots<k_n$. 
For the half-size pairing scheme, we choose $k_1=k_i$ and $k_2=k_{i+n/2}$; for the head-tail pairing scheme, we choose $k_1=k_{i}$ and $k_{2}=k_{n-i}$. In numerical simulation, for each fix $k_{\text{max}}$ and $N$, we random sample $k_{\text{true}}\in [0,k_{\text{max}}]$, and follow predictive dissonance protocol to find $k_{\text{true}}$. And we use ensemble averaged $\overline{T}$ to denote the average running time to pinpoint $k_{\text{true}}$ for given $N$ and $k_{\text{max}}$. For those two pairing schemes, we first fix $k_{\text{max}}=50$, and then vary the number of items $N$ in the database. The result is shown in Fig.\ref{fig:scaling}(a). We find for both pairing schemes, $\overline{T}\propto \sqrt{N}$. This is reasonable, because each run time is proportional to $\sqrt{N}$ whichever pairing scheme is chosen. Therefore, ensemble averaged run time should also be proportional to $\sqrt{N}$. 

{}

Next we study the relation between averaged run time $\overline{T}$ and $k_{\text{max}}$. We fix $N=20000$ and vary $k_{\text{max}}$. As shown in Fig.\ref{fig:scaling}(b), we find that $\overline{T}\propto k_{\text{max}}^{\alpha}$ with 
the power $\alpha$ depending on the pairing scheme. For the half-size pairing, $\overline{T}\propto k_{\text{max}}^{0.59}$, while for head-tail pairing, $\overline{T}\propto k_{\text{max}}^{0.68}$. We conjecture that the lower bound for $\alpha$ is 0.5, because we can roughly estimate that $\overline{T}\propto k_{\text{max}}\sqrt{N/k_{\text{max}}}\propto \sqrt{k_{\text{max}}N}$. What pairing scheme can achieve the optimal lower bound is subject to further discussion. There are problems where the number of solutions
$k_{\rm true}$ scale with $N$~\cite{WYW}. If $k_{\rm true}\propto N^\gamma$, our numerical results indicate that
$\overline{T}\propto N^{\alpha\gamma+0.5}$.

\section{Robust Readout} 

We now briefly describe a very different way to exploit monitor qubits to address the same problem.  It is conceptually simpler and potentially much faster, but it requires additional resources and it depends upon assumed physical properties of qubits.    Indeed, let us assume that we have an ensemble containing several monitor qubits, each of the kind described before, and that they are localized particles - ``spins'' - carrying a magnetic moment, all within a common small region.   Then the systematic oscillation of the ensemble of monitor monitor bits will set up an oscillating magnetic field, which can be read out with great sensitivity, for instance using a SQUID.  The frequency of that oscillating field encodes the unknown value of $k$, according to our preceding formulae.    Use of several monitor qubits, of course, also brings in protection against errors in any one of them, and against small uncorrelated errors that affect all of them.  




\section{Summary} We have used resonance to construct quantum search algorithms. 
We have shown how to add monitor qubits that check for resonance without 
disturbing the computational qubits. 
One can use monitor qubits to implement predictive dissonance and robust readout, which allow us to find the number of answers efficiently when that is unknown.  

Our algorithms illustrate the importance of physical considerations in assessing computational potential.  The parameter $p$, which governs overall speed, represents interaction energy at a particular frequency, and could become quite large in a resonant context.  Robust readout can in principle obviate $k$ dependence altogether.  We indicated in broad terms how robutst readout can be implemented using spin qubits.  Both this and possible alternative implementations merit further study. 

B.W. is  supported by the The National Key R\&D Program of China (Grants No.~2017YFA0303302, No.~2018YFA0305602). F.W. is 
supported by the Swedish Research Council under Contract No. 335-2014-7424, 
U.S. Department of Energy under grant Contract No.de-sc0012567,  and by the European Research Council under grant 742104.

\bibliography{qalgorithm}

\end{document}